\documentclass[prl,amsmath,amssymb, twocolumn, superscriptaddress]{revtex4-1}
\usepackage{graphicx}
\usepackage{dcolumn}
\usepackage{bm}
\usepackage{rotating}
\usepackage{siunitx}
\usepackage{multirow}
\usepackage{booktabs}
\usepackage{afterpage}
\usepackage[table]{xcolor}
\usepackage[english]{babel}
\usepackage{amssymb} 
\usepackage{amsmath} 
\usepackage[utf8]{inputenc} 
\usepackage{longtable}
\usepackage{placeins}

\newcommand{\be}{\begin{equation}}
\newcommand{\ee}{\end{equation}}
\newcommand{\bea}{\begin{eqnarray}}
\newcommand{\eea}{\end{eqnarray}}

\def\a{\alpha}

\def\G{\Gamma}
\def\d{\delta}

\begin{document}
\widetext 
\title{Critical Role of the Exchange Interaction for the Electronic Structure and Charge-Density-Wave Formation in TiSe$_2$}
\author{Maria Hellgren}
\affiliation{Institut de Min\'eralogie, de Physique des Mat\'eriaux et de Cosmochimie, Sorbonne Universit\'es, Universit\'e Pierre et Marie Curie, CNRS, IRD, MNHN, 4 Place Jussieu, 75252 Paris, France}
\affiliation{Physics and Materials Science Research Unit, University of Luxembourg, 162a avenue de la Fa\"{i}encerie, L-1511 Luxembourg, Luxembourg}
\author{Jacopo Baima}
\affiliation{Institut de Min\'eralogie, de Physique des Mat\'eriaux et de Cosmochimie, Sorbonne Universit\'es, Universit\'e Pierre et Marie Curie, CNRS, IRD, MNHN, 4 Place Jussieu, 75252 Paris, France}
\author{Raffaello Bianco}
\affiliation{Dipartimento di Fisica, Universit\`{a} di Roma La Sapienza, Piazzale Aldo Moro 5, I-00185 Roma, Italy}
\affiliation{Institut de Min\'eralogie, de Physique des Mat\'eriaux et de Cosmochimie, Sorbonne Universit\'es, Universit\'e Pierre et Marie Curie, CNRS, IRD, MNHN, 4 Place Jussieu, 75252 Paris, France}
\author{Matteo Calandra}
\affiliation{Institut de Min\'eralogie, de Physique des Mat\'eriaux et de Cosmochimie, Sorbonne Universit\'es, Universit\'e Pierre et Marie Curie, CNRS, IRD, MNHN, 4 Place Jussieu, 75252 Paris, France}
\author{Francesco Mauri}
\affiliation{Dipartimento di Fisica, Universit\`{a} di Roma La Sapienza, Piazzale Aldo Moro 5, I-00185 Roma, Italy}
\affiliation{Graphene Labs, Fondazione Istituto Italiano di Tecnologia, Via Morego, I-16163 Genova, Italy}
\author{Ludger Wirtz}
\affiliation{Physics and Materials Science Research Unit, University of Luxembourg, 162a avenue de la Fa\"{\i}encerie, L-1511 Luxembourg, Luxembourg}
\date{\today}
\pacs{}
\begin{abstract}
We show that the inclusion of screened exchange via hybrid functionals provides a unified description of the electronic and vibrational properties of TiSe$_2$. In contrast to local approximations in density functional theory, the explicit inclusion of exact, non-local exchange captures the effects of the electron-electron interaction needed to both separate the Ti-$d$ states from the Se-$p$ states and stabilize the charge-density-wave (CDW) (or low-T) phase through the formation of a $p-d$ hybridized state. 
We further show that this leads to an enhanced electron-phonon coupling that can drive the transition even if a small gap opens 
in the high-T phase. Finally, we demonstrate that the hybrid functionals can generate a CDW phase where the electronic bands, 
the geometry, and the phonon frequencies are in agreement with experiments.
\end{abstract}
\keywords{}
\maketitle
The charge density wave (CDW) instability is a common phenomenon in layered 
semi-metallic transition metal dichalcogenides (TMDs) \cite{wilson_charge-density_1975} 
and has attracted considerable interest over the years, from both the experimental and 
the theoretical side. The CDW phase is often found to compete with superconductivity 
and thus plays a similar role as the anti-ferromagnetic phase in strongly correlated 
heavy fermion systems or in high-T$_{\rm c}$ cuprates \cite{PhysRevLett.99.107001,kuspress09,morosancu06,hasan}. 
This intriguing similarity has stimulated the search for a better understanding of the 
physical mechanism behind the CDW instability in TMDs \cite{rossnagel_origin_2011}. 

The CDW instability in TiSe$_2$ is one of the most studied and debated. 
On the experimental side, neutron diffraction \cite{diSalvo76} and X-ray scattering \cite{abbamonte} 
have established the existence of a commensurate $2\times2\times2$ structural transition at 200 K. 
This is confirmed by angle resolved photo emission spectroscopy (ARPES) as well as by transport 
measurements \cite{diSalvo76}, where an abrupt increase in resistivity is found at the same 
temperature.  However, upon further cooling the resistivity reaches an anomalous maximum
\cite{diSalvo76}, after which a weak metallic behaviour is observed. By contrast, ARPES 
finds an insulating low-T phase, with a gap of approximately 0.15 eV \cite{Pillo2000,kidd02,ross02,monney07,ChenChiang2015,ChenChiang2016}. In the high-T phase 
ARPES has not been able to conclude whether the  system is semi-metallic or semi-conducting due 
to the very small indirect (possibly negative) gap. Theoretically, this fact 
makes TiSe$_2$ an ideal candidate to exhibit an excitonic insulator phase \cite{kohn_exc,Pillo2000,monney09,monney11} 
for which the CDW transition is driven by a purely electronic instability. Some recent 
experiments \cite{rohwer} partly support this scenario. On the other hand, excitonic correlations 
alone are insufficient as demonstrated in Ref. \onlinecite{porer}. 

Additional experimental evidence for the CDW instability has been provided by
vibrational spectra as a function of temperature. A complete softening of an 
optical phonon at the $L$-point has been observed in inelastic X-ray scattering 
experiments \cite{weber11}. In Raman and infrared (IR) spectroscopy 
the transition is detected by the appearance of a large number of new modes \cite{Holy,Snow03}, 
some of which can be related to the CDW transition due to their strong temperature dependence. 

On the theoretical side, density functional theory (DFT) within the standard 
local density approximation (LDA) or the semi-local PBE approximation predicts a 
structural instability at the correct wave vector \cite{calandra11,weber11,bianco2015} 
emphasising the role of the lattice distortion. However, orbital occupations obtained from the 
electronic band structure are in disagreement with ARPES measurements and CDW phonon 
frequencies strongly deviate from experiments. This suggests that a proper inclusion of 
exchange and correlations effects may be crucial to describe the CDW instability. It has been shown that 
the inclusion of nonlocal exchange already gives a much better description of the electronic 
bands in both the high- and low-T phase \cite{Cazzaniga2012,vydrova2015,ChenChiang2015,ChenChiang2016,udo2012}. 
In Ref. \onlinecite{bianco2015} it was further shown that also the DFT+U approach improves 
the electronic band structure, bringing it into good agreement with ARPES spectra. 
This seeming improvement, however, was accompanied by a complete loss of the CDW instability.

In this work we provide a unified description of both the electronic band structure and the 
lattice dynamics of TiSe$_2$, and give a physical explanation for the CDW instability based 
on first-principle calculations. By an inclusion of exact-exchange within the hybrid functionals 
we capture the strong Coulomb repulsion due to the localised Ti-$d$-states, thus correcting the 
orbital occupations. We also capture the long-range exchange interaction which enhances the 
electron-phonon coupling (EPC) and stabilizes the CDW phase through the formation of a 
Ti-$d$-Se-$p$ hybridized state. 
\begin{figure}[t]
\includegraphics[width=\columnwidth]{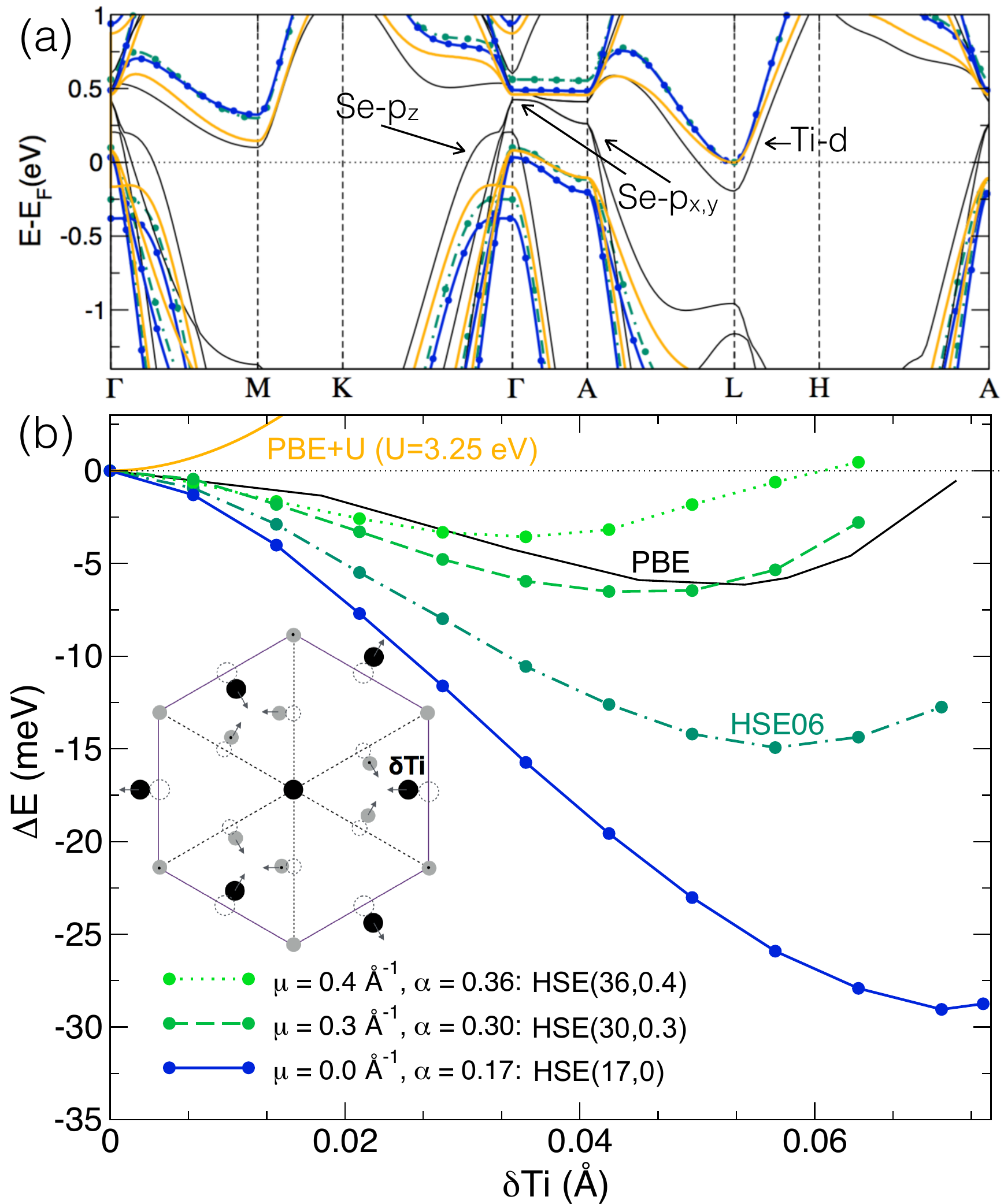}
\caption{The upper panel (a) shows the band structure in the high-T phase in PBE (black solid line), 
HSE06 (green dash-dotted line and circles), HSE(17,0) (blue solid line and circles) and PBE+U (U=3.5 eV) 
(orange solid line). The lower panel (b) shows the corresponding energy gain as a 
function of $\d$Ti. We show also the results obtained by varying the $\alpha$ and $\mu$ parameters 
such that the band structure around conduction band minimum and valence band maximum is the same 
as the one of HSE06 and PBE+U.}\label{e_gain}
\end{figure}

TiSe$_2$ is a layered material which in the high-T phase ($>$200 K) 
crystallises in the space group $P\bar 3m1$. The CDW phase 
is characterised by a 2$\times$2$\times$2 superstructure with a different 
space group $P\bar 3c1$. The PBE phonon dispersion of the high-T phase was 
analysed in Refs. \onlinecite{calandra11} and \onlinecite{bianco2015}. The distortion 
pattern ${\bf d}_{3L}$ (see Eq. 9 in Ref. \onlinecite{bianco2015}) associated with the 
imaginary phonon frequencies at the three 
equivalent $L$ points is of A$_u$ symmetry and its projection on a single TiSe$_2$ layer is shown 
in the inset of Fig.~\ref{e_gain}(b).  It is completely identified by its symmetry, the
magnitude of the displacement $\delta {\rm Ti}$ and the ratio 
$\delta {\rm Ti}/\delta {\rm Se}$. The layers are held together 
by van der Waals (vdW) forces and in the CDW phase adjacent layers are rotated 
by 60 degrees. 
We verified that the vdW forces \cite{grimme} do not play a role in the CDW distortion and by using the experimentally 
determined $a$ and $c$ lattice parameters \cite{RIEKEL1976structure}, we indirectly account for the vdW contribution to the structure 
(see also Refs. \cite{bianco2015,SuppMat}). Although the spin-orbit coupling (SOC) splits the Se-$p$ energy 
levels \cite{vydrova2015} it only gives a small contribution to the energetics \cite{SuppMat}. 
Due to the increased computational cost we have therefore omitted SOC in the calculations presented here. 
Local functionals in DFT such as PBE give rise to an excess of electron occupation 
on the Ti-$d$-states compared to ARPES, mainly due to the large $p_z-d$ band overlap 
but also due to the overestimated $p_{x,y}-d$ band overlap (see black curve in Fig.~\ref{e_gain} (a)). 
To cure this problem it is necessary to include the Coulomb repulsion or 'U' due to the 
localised Ti-$d$ states. A simple corrective procedure to 
incorporate this effect is the DFT+U approach \cite{LDAU}. Indeed, in Ref. \onlinecite{bianco2015} 
it was shown that the $p-d$ band overlap reduced significantly (see orange curve Fig.~\ref{e_gain} (a)). 
However, it was also found that the energy gain in the distortion decreases by increasing 'U' \cite{SuppMat}.  
The hybrid functionals contain a fraction of exact-exchange and therefore naturally incorporate the
repulsive 'U' interaction \cite{degironc}. In addition, they exhibit the long-range exchange interaction, 
which generates the attractive electron-hole interaction at the linear response level \cite{kressehse,rerat}. 
The description of screening is rather simplified using a pre-factor $\a$ and a range-separated Coulomb 
potential defined by $\mu$. These parameters can 
be fitted in various ways but often the HSE06 parameters ($\a=0.25,\mu=0.2 \,{\rm \AA}^{-1}$) give good 
results \cite {hse03}. In this work we have used the freedom of varying ($\a,\mu$) to reveal the 
critical role played by exchange, and to shed light on the mechanism that drives the CDW 
transition in TiSe$_2$. 
\begin{figure*}[t]
\includegraphics[width=2\columnwidth]{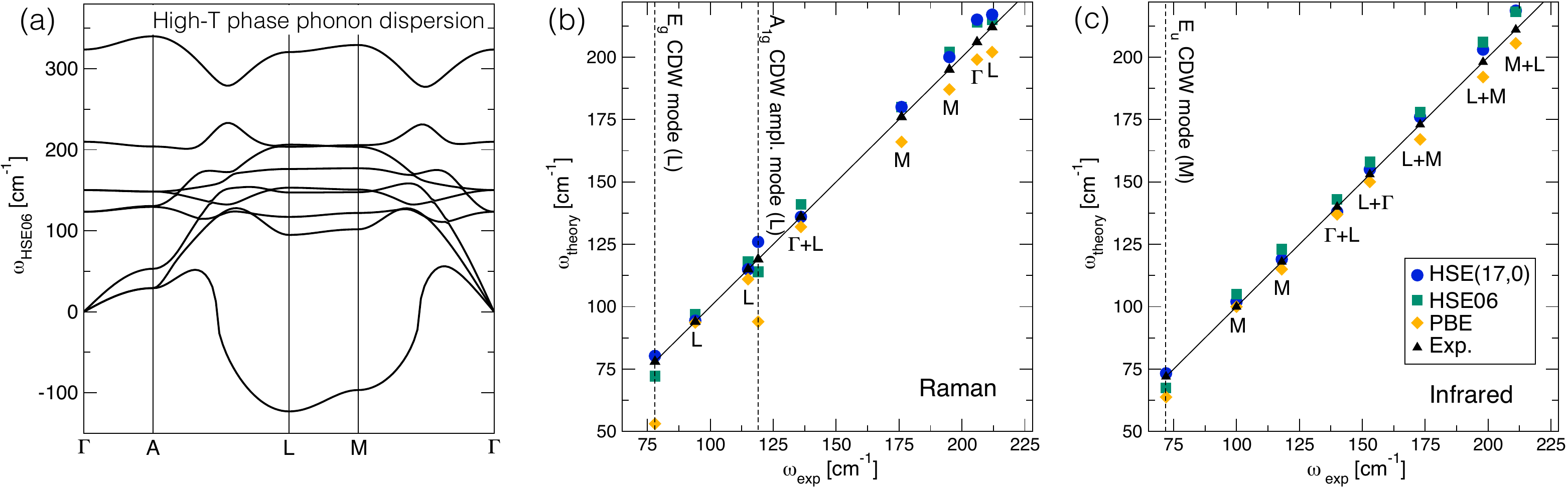}
\caption{(a) Phonon dispersion of the high-T phase (HSE06). 
(b-c) Experimental \cite{Sugai,Holy} versus calculated phonon frequencies of Raman and IR modes 
in the low-T phase. Modes related to the CDW distortion are marked by vertical dashed lines. 
The labels refer to the largest components of the phonon eigenvector with respect to the 
eigenvectors of the high-T structure.}\label{phononfig}
\end{figure*}

In Fig.~\ref{e_gain} (a) we plot the results for the band structure in the high-T phase
with PBE, PBE+U and HSE06 \cite{SuppMat,qe,crystal14,vasp1,vasp2,vasp3}. 
In Fig.~\ref{e_gain} (b) we plot the corresponding energy gain as a function of $\d{\rm Ti}$ 
by displacing the atoms according to the ${\bf d}_{3L}$ 
CDW pattern (using the experimentally determined ratio $\delta {\rm Ti}/\delta {\rm Se}=3$ \cite{diSalvo76}). 
We found HSE06 to give an enhanced negative curvature and energy gain of approximately a 
factor of 2 compared to PBE. This difference is mainly attributed to an overscreening 
at the PBE level, i.e., due to the overestimation of the metallicity of the system. 
However, this is not the complete picture.
We see that PBE+U and HSE06 have almost identical orbital occupations and gaps - 
both in rather good agreement with experiment \cite{ChenChiang2015,ChenChiang2016,SuppMat}  
- but in PBE+U the instability is absent. This suggest that 
the long-range (or nonlocal) Coulomb exchange interaction is important for the formation
of the CDW phase, as demonstrated explicitly in Figure~\ref{e_gain} (b). 
We have calculated the energy gain using a set of ($\a,\mu$) parameters 
all tuned to give similar band-structures to HSE06 and PBE+U. By reducing the range, i.e., by increasing $\mu$, 
thus approaching the PBE+U limit, the instability quickly vanishes (green curves). Instead, with an infinite 
range (HSE(17,0)) the energy gain increases substantially as does the amplitude of the distortion. 
After relaxing the atomic positions in the CDW phase, 
we find $\d {\rm Ti} = 0.055,0.061,0.082 \,{\rm \AA} $ with PBE, 
HSE06 and HSE(17,0) respectively. In experiment, $\d {\rm Ti} = 0.085 \pm 0.014 \,{\rm \AA}$ \cite{diSalvo76}, 
thus suggesting a long-range (or unscreened) Coulomb interaction. 
Based on the above analysis, we will continue to discuss the mechanism for the CDW transition, but first we analyze the 
properties of the CDW phase.

The electronic band structure in comparison with ARPES suggests that the CDW phase is accurately 
captured with hybrid functionals \cite{SuppMat,ChenChiang2016}. 
However, as seen above, a good description of the band structure alone is insufficient. 
The vibrational frequencies are proportional to 
the second derivative of the energy and thus provide an additional stringent test for the validity 
of different approximations. We have calculated the harmonic Raman and IR active phonon 
frequencies in the CDW phase within both HSE06 and the infinite range HSE(17,0) using a finite 
difference approach as implemented in the {\tt CRYSTAL} code \cite{vibCRYS}. 

In Fig.~\ref{phononfig} we compare the theoretical frequencies with low-T experimental Raman 
\cite{Sugai} and IR \cite{Holy} frequencies (at 11 K and 20 K, respectively). We analyze the 
spectra of the CDW phase by back-folding of the phonons at $A$, $L$ and $M$ of the high-T phase 
Brillouin zone (see HSE06 phonon dispersion \footnote{The phonon dispersion is approximated on 
a 2$\times$2$\times$2 q-grid.} in Fig.~\ref{phononfig} (a)) onto the $\bar\Gamma$ point 
of the low-T phase Brillouin zone \cite{SuppMat}. In the following, the 
high-symmetry points of the 2$\times$2$\times$2 supercell are marked with a bar.
For all three approximations, we find an overall good agreement with the experimental results \cite{uchida1981IR_Ram,Sugai,Holy,Liang,wilson1978IR_rev,duong2017raman}. However, 
there are three notable exceptions of particular interest for this work: 
these are the modes related to the instabilities of the high-T structure (marked by vertical dashed lines). 
The three-fold degenerate imaginary phonon frequency at $L$ splits into one A$_{1g}$ and a two-fold 
degenerate E$_g$ mode, which are both Raman active. 
These have been previously identified as 
being related to the CDW transition from their experimental temperature dependence \cite{Holy,Snow03}. 
In fact, the A$_{1g}$ (amplitude) mode corresponds to the oscillations of the CDW order parameter. 
The imaginary phonon frequency at $M$ similarly gives rise to an E$_u$ IR active phonon and to 
one A$_{1u}$ inactive mode. The frequencies of the CDW modes are systematically underestimated 
by PBE (20\% and 30\% for the A$_{1g}$ and E$_g$ modes, respectively). HSE06 brings these 
frequencies in much better agreement with experiment (now underestimated by 4\% and 7\%, respectively). 
HSE(17,0), which gave the best geometry, also gives excellent 
CDW mode frequencies (overestimated by 5\% and 2\%, respectively). Since the frequencies
of these modes depend very sensitively on the CDW potential energy surface (compare Fig.~\ref{e_gain} (b)), 
they represent an important confirmation that the hybrid functionals reliably reproduce the electronic 
structure of TiSe$_2$.

\begin{figure}[t]
\includegraphics[width=\columnwidth]{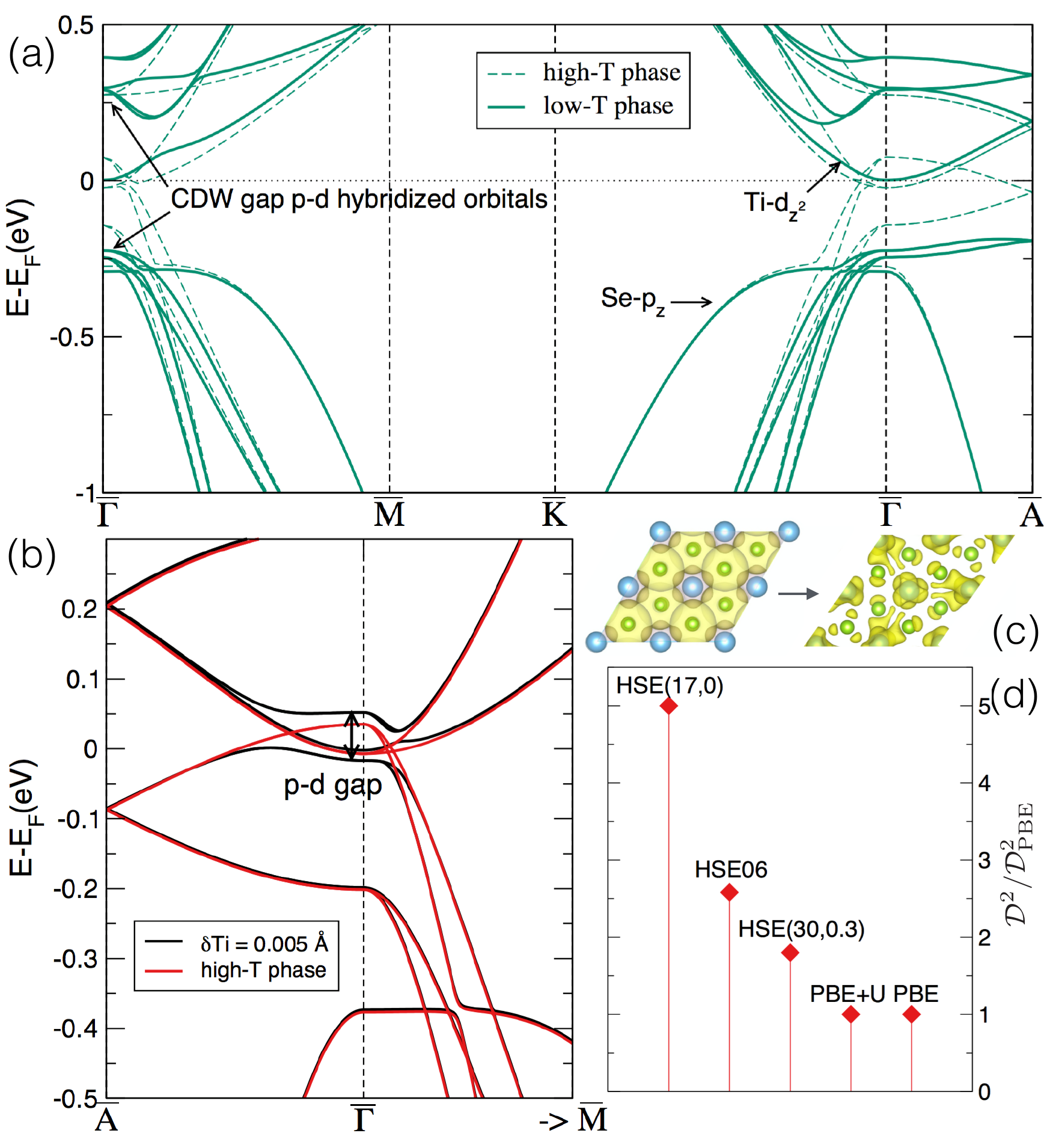}
\caption{(a) HSE06 band-structure of undistorted (high-T) and CDW (low-T) phases of TiSe$_2$. Panel (b) shows a zoom 
around the Fermi level and the EPC between the $p-d$ states is calculated in panel (d)).
The value for ${\cal D}^2_{\rm PBE} = 8 \,{\rm eV}^2/{\rm \AA}^2$ (where $d_{3L}$ is measured in units of $\d$Ti at constant $\delta {\rm Ti}/\delta {\rm Se}=3$). 
Panel (c) demonstrates the mixing (hybridization) of these orbitals \cite{vesta} in the CDW phase.
}
\label{electronphonfig}
\end{figure}

With a reliable description of the vibrational and electronic properties, we can now analyze 
in detail the physical mechanism that is responsible for the CDW instability in TiSe$_2$. 
 Figure~\ref{electronphonfig} (a) shows the HSE06 band structure 
in both the low-T and high-T phase (the latter one is back-folded into the Brillouin zone that 
corresponds to the 2$\times$2$\times$2 supercell of the low-T phase). At the CDW distortion, the three-fold 
degenerate Ti-$d$ band splits into a band with dominant $d_{z^2}$ character derived from the Ti 
atom in the supercell that does not move with the distortion, and a two-fold degenerate band with 
Ti-$d$ and Se-$p$ hybridization in a small region around $\bar\Gamma$ due to the interaction between the $p$ and $d$ states. 
Similarly, the Se-$p$ bands move to lower energy and hybridize with Ti-$d$ states 
around $\bar\Gamma$. Such $p-d$ hybridization, discussed also in Refs. \onlinecite{PhysRevB.31.8049,ross02,wezel}, 
is observed by studying the site-projected orbitals onto spherical harmonics before and after the 
transition. Panel (b) presents a zoom of the 
region around $\bar\G$, showing the effect of a small ${\bf  d}_{3L} $ distortion on the 
electronic structure. We see that the distortion only changes the $p-d$ gap while leaving the rest of
the electronic structure almost unaffected. We can therefore assume that it is mainly the $p-d$ states around $\bar\G$
which are involved in the transition and extract the EPC, or, more precisely, 
the deformation potential $\cal D$ for the different approximations. The deformation potential is defined as 
${\cal D}=\langle{\bar\G d}|\d V/\d d_{3L}|\bar\G p\rangle$, where the variation of the total 
self-consistent potential $V$ is taken with respect to the magnitude $d_{3L}$ of the ${\bf d}_{3L}$ distortion. 
The states $|\bar \G d\rangle$ and 
$|\bar\G p\rangle$ with energy difference ${\cal E}_0$ are obtained from the Ti-$d$ and Se-$p$ states 
at $L$ and $\Gamma$ in the undistorted unit cell by back-folding them onto $\bar\Gamma$ in 
the supercell. Without loss of generality, all quantities can be taken as real in the supercell. 
The $p$ and $d$ states are superimposed and illustrated in 
Fig. \ref{electronphonfig} (c) together with the $p-d$ hybridized state in the CDW phase. 
We also note from first order perturbation theory that 
${\cal D}=\langle{\delta(\bar\G d)/\d{d}_{3L}}|\bar\G p\rangle\times {\cal E}_0$ and hence 
(at fixed ${\cal E}_0$) ${\cal D}$ measures the overlap between the variation of the $d$ 
state with the distortion and the undistorted $p$ state.
We then obtain $\cal D$ using a finite difference approach \cite{PhysRevB.78.081406}. At small 
but finite ${d}_{3L}$ we construct the $2\times2$ submatrix of the Hamiltonian $\cal H$: ${\cal H}_{dd}=-{\cal E}_0/2+{{\cal O}({d}_{3L}^2)}$, ${\cal H}_{pp}={\cal E}_0/2+{{\cal O}({d}_{3L}^2)}$ and 
${\cal H}_{pd}={\cal H}_{dp}={\cal D}{ d}_{3L}+{{\cal O}({d}_{3L}^2)}$ and diagonalize it to obtain ${\cal D}^2=\frac{1}{4}({\cal E}^2-{\cal E}_0^2)/{d}_{3L}^2$, where ${\cal E}$ is the $p-d$ gap upon the distortion and we have discarded 
the higher order terms in $d_{3L}$. The results obtained from this procedure 
can be found in Fig.~\ref{electronphonfig} (d). Although some coupling already exists 
between $p$ and $d$ states at the PBE level, the deformation potential (and so the EPC) 
is strongly enhanced via the nonlocal exchange potential in $V$ provided by the inclusion 
of long-range exchange. Indeed, a comparison between PBE and PBE+U shows that U has no effect on the EPC. 
For the approximations with the same band structure we see the same trend in Figs. \ref{e_gain} (b) 
and \ref{electronphonfig} (d) and hence we conclude that it is the nonlocal exchange interaction that 
determines the strength of the instability. The inclusion of SOC splits and slightly changes the 
dispersion of the Se-$p$ bands close to the Fermi level but has only a small 
quantitative effect on the EPC \cite{soc,SuppMat}.
Our result goes in line with some previous findings that the EPC can be strongly enhanced by nonlocal 
exchange, notably in weakly doped 2D materials \cite{PhysRevB.78.081406,doi:10.1021/nl9034626,
PhysRevLett.114.077001,PhysRevB.94.035101} and some high-T$_c$ superconductors \cite{PhysRevX.3.021011}. 
Although the band structure in HSE06, which we used as a 
reference, has a small negative $p-d$ gap (-0.1 eV) in the high-T phase, the instability does not 
crucially depend on the existence of a Fermi surface. In fact, due to the enhanced EPC, the CDW phase 
can exist even if the gap in the high-T phase is positive. This is easily demonstrated by increasing 
the amount of exact-exchange ($\a$). With $\mu=0.2\, {\rm \AA}^{-1}$ fixed, the instability persists 
up to $\a=0.35$ when the gap is as large as 0.2 eV. At larger values of the gap 
(or $\a$) the hybridization of Ti-$d$ -and Se-$p_{x,y}$ bands is suppressed and the instability 
disappears. If we instead decrease $\a<0.25$ the energy gain reaches a maximum around $\a=0.15$. At smaller
values of $\a$ the CDW phase becomes metallic and the energy gain drops. This unusual 
behaviour as a function of $\a$ (or the gap) shares some features with the excitonic insulator 
transition (EIT) of Kohn et al.\cite{kohn_exc,kohn2}. 
The EIT arises in a two band model (similar to the $p$ and $d$ bands in TiSe$_2$) when tuned 
until the value of the $p-d$ gap is smaller than the exciton binding energy. As a result, 
$p$ and $d$ orbitals mix and form a new ground state with lower symmetry. Since the symmetry of the lattice is kept 
fixed, the EIT is a purely electronic effect. In our calculations it was not possible to generate 
a symmetry lowering of the electronic density alone. The mechanism that we found relies on the simultaneous 
lattice distortion and strong EPC, enhanced by the exchange interaction. 
Whether the system is still influenced by excitonic effects can be determined by studying the 
excitation spectrum, but this we leave for future work.



M.H. and L.W. gratefully acknowledge financial support from the FNR (projects RPA-phonon and INTER/ANR/13/20/NANOTMD). 
The authors acknowledge support from the European Union Horizon 2020 research and innovation programme under 
Grant agreement No. 696656-GrapheneCore1. and by Agence Nationale de la Recherche under the reference
n. ANR-13-IS10-0003-01 and by Prace. Calculations
were performed at IDRIS, CINES, CEA and BSC TGCC. The authors are grateful to K. Rossnagel, for having supplied 
to them the image data of ARPES experiment.


%
\cleardoublepage 

\begin{widetext}
\section{Supplemental material for: \\
Critical Role of the Exchange Interaction for the Electronic Structure and Charge-Density-Wave Formation in TiSe$_2$}
Below we present further details on the computational procedures, high-T phase crystal structure, phonon frequencies, 
PBE+U calculations, spin-orbit coupling and electronic structure compared to ARPES.
\section{Computational details}  
Results with PBE and PBE+U were obtained with Quantum ESPRESSO \cite{qe} and results with HSE06 
with VASP \cite{vasp1,vasp2} within the projector-augmented-wave (PAW) method. \cite{vasp3} A 
plane-wave cut-off of 320 eV was used and semi-core electrons were included in the Ti PAW potential. Results were 
converged with a $24\times 24\times 12$ $k$-point mesh in the high-T cell. For most calculations presented in the
manuscript we used the all-electron {\tt CRYSTAL} code \cite{crystal14} with an adapted molecular def2-TZVP basis 
set for solid state calculations. The results for the energy gain in HSE06 was within 1 meV/supercell in agreement 
between VASP and {\tt CRYSTAL} (see Fig.~\ref{socnosoc} below). 
\section{High-T phase crystal structure}
In Table \ref{tab:sez-highT} we present the experimental and theoretical lattice parameters and Se-$z$-positions 
of the high-T phase structure. The parameter $a$ is the in-plane hexagonal lattice constant, $c$ is the distance between layers 
and $h$ is the distance between the Ti and Se layers. The hybrid functionals in combination with a simple 
Grimme D2\cite{grimme} van der Waals correction give lattice parameters in good agreement with experiments. 
In our calculations we have, however, consistently used the experimental lattice parameters but relaxed 
atomic positions. Only the energy gain curves in Fig.~1 (b) of the manuscript were obtained with fixed Se-$z$-positions and fixed 
$\delta {\rm Ti}/\delta {\rm Se}(=3)$ ratio. At the minimum the atomic positions were again relaxed as described 
in the manuscript. In Fig. 1 we show the in-plane crystal structure in the high -and low-T phases. Big blue dots signifies Ti atoms and small green dots Se atoms. A dot on the Se atom indicates that the atom is on the opposite side of the Ti layer as compared to the Se atom without the dot. We have also replotted in a larger scale the $p$, $d$ and $p-d$-hybridized orbitals of Fig. 3 (c) in the manuscript.

\begin{table}[h!]
\caption{Lattice parameters are calculated within PBE, HSE06 and HSE(17,0). The van der Waals forces are accounted for via a Grimme D2\cite{grimme} correction.}
\label{tab:sez-highT}
\begin{tabular}{c c c c c}
\hline
  & Exp.\cite{RIEKEL1976structure} & PBE  & HSE06 & HSE(17,0) \nonumber \\
a(${\rm \AA}$) & 3.540 &  3.519  & 3.528  & 3.531 \nonumber \\
c(${\rm \AA}$) & 6.007 &  6.130  & 6.104  & 6.113 \nonumber \\
h(${\rm \AA}$) & 1.532 &  1.555  & 1.522   &  1.527  \nonumber \\
\hline
\end{tabular}
\end{table}
 \begin{figure}[b]
\includegraphics[scale=0.25]{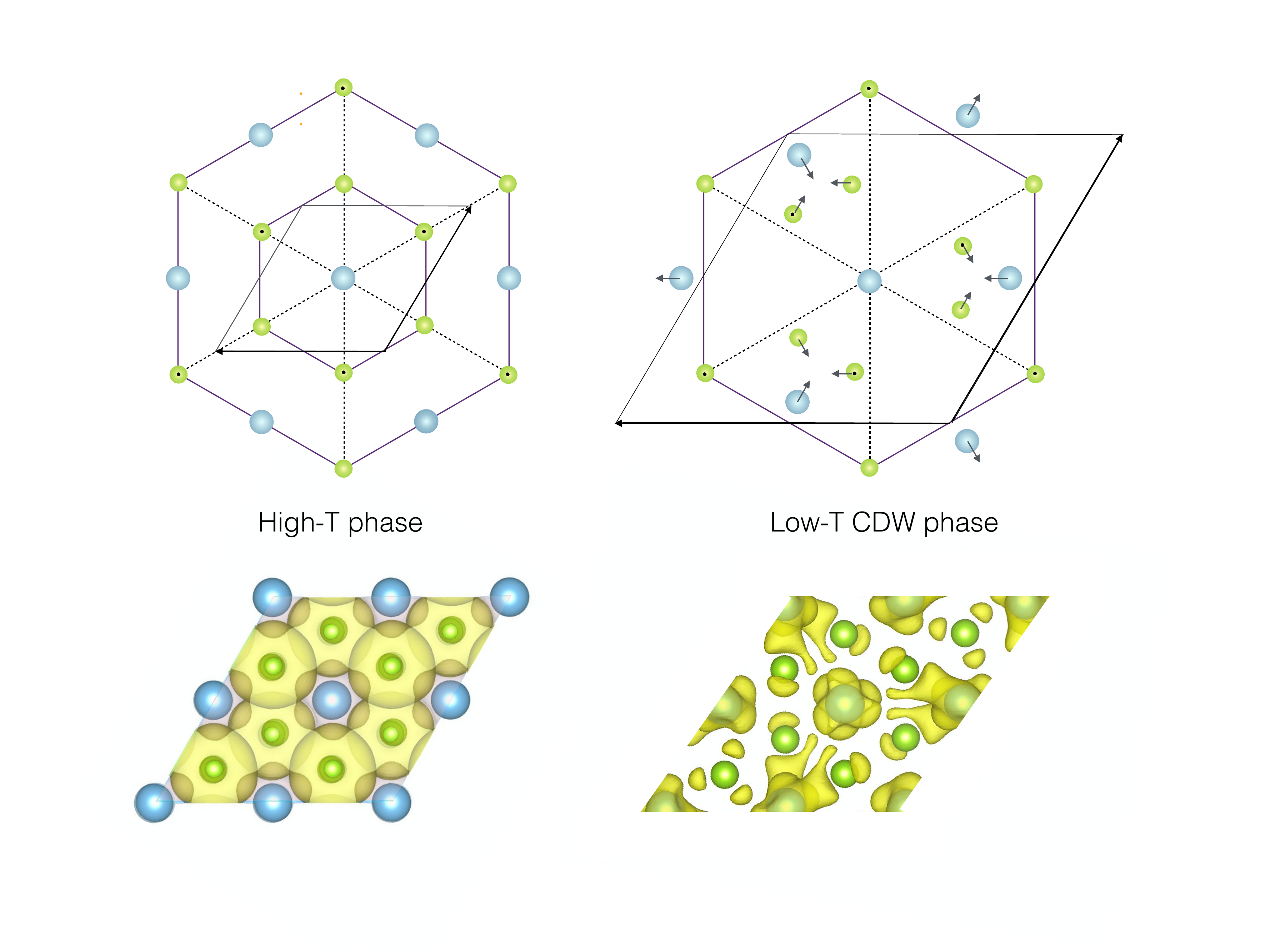}
\caption{In-plane crystal structure of the high -and low-T phases and the corresponding $p$, $d$ and $p-d$-hybridized orbitals that participate in the CDW transition.}
\end{figure}

\section{Phonon frequencies in the high-T phase}
The phonon frequencies in the high-T phase are reported in Table \ref{tab:ph-highT}.
In the high-T phase the phonons are measured at room temperature and thus cannot be calculated exactly within the harmonic approximation. Indeed, the two Raman active phonons at 200 cm$^{-1}$(A$_{1g}$) and at 136 cm$^{-1}$ (E$_g$) are overestimated by hybrid functionals. It is expected that thermal effects would soften these values, and this trend was seen in experiment at least for the E$_g$ mode in the high-T phase.\cite{duong2017raman} On the contrary, the IR active mode around 140 cm$^{-1}$ is underestimated, with both HSE06 and HSE(17,0). This mode belongs to the same phonon branch as the soft mode at $L$ which drives the instability and should thus harden significantly with temperature. Such behaviour can also be seen in experiment.\cite{uchida1981IR_Ram}

\begin{table}[h!]
\caption{Calculated zone center phonon frequencies (cm$^{-1}$) for the high-T structure compared with
Raman (R) and Infrared (I) experiments.}\label{tab:ph-highT}
\hspace*{-3cm}\begin{tabular}{c c c c c}
\hline
Symm. & Exp. &  PBE   &   HSE06  & HSE(17,0)  \\
E$_g$  (R) &  136\cite{Sugai} (273 K)  / 134\cite{Holy} (300 K)   &    136.9    &    147.5  & 147.3  \\
E$_u$ (I) &  143\cite{Liang} (300 K) /137\cite{Holy} (300 K)         &    141.5      &   124.8  & 101.1  \\
A$_{1g}$ (R) &    200\cite{Sugai} (273 K) /194\cite{Holy} (300 K)   &    195.6     &   209.5  & 207.9  \\
A$_{2u}$        &     -                      &    292.7      &   323.8  & 314.4  \\
\hline
\end{tabular}\hspace*{-3cm}
\end{table}

\newpage
\section{Phonon frequencies in the low-T phase}
Tables \ref{tab:Ram_lowT}, \ref{tab:IRlowT} and \ref{tab:IRdist_und} report the computed Raman and infrared active frequencies in the CDW phase, comparing them to the peaks identified in several experimental spectra.
They also contain an analysis of the CDW spectra in terms of back-folding of phonon modes from the Brillouin zone of the high-T phase.
The $q$-points of the original lattice folded onto $\bar\Gamma$ in the supercell
are $\Gamma$ and $A$ with weights 1, plus $M$ and $L$ with weights 3. By applying crystalline translations, all the phonon eigenvectors 
at $\bar\Gamma$ of the undistorted supercell are obtained from the phonon eigenvectors 
at the previously mentioned $q$-points of the unit cell.
The degeneracy between frequencies at symmetry equivalent points is broken by the distortion, so that, e.g., 
the three degenerate modes from $L_1$, $L_2$ and $L_3$ split into an $E$ and an $A$ representation.

We label  $|\sigma_{\bf{q}}\rangle$ the phonon eigenvectors at $\bar\Gamma$ of the high-T phase. $\bf{q}$ labels the $q$-point the mode is back-folded from. 
We then label $|\bar{\nu}\rangle$ the eigenvectors of the distorted low-T phase.
We define the decomposition of a mode $|\bar{\nu}\rangle$ as the square overlap with the eigenvectors of the undistorted supercell $|\langle \bar{\nu}|\sigma_{\bf{q}}\rangle|^2$, and then sum over degenerate frequencies in the undistorted supercell. For any given $|\bar{\nu}\rangle$ we report in the tables the largest square overlaps as a function of $\sigma$ (the ones larger than 10\%), the frequency $\omega_\sigma$ and the point $\bf{q}$.




\begin{sidewaystable}[h!]
  \caption{Raman peaks identified in experimental spectra of the CDW phase (at 53 K and 11K) compared to computed
    Raman active frequencies (cm$^{-1}$). The point(s) {\bf q} in the Brillouin zone and the frequency(s)
    $\omega_{\rm undist.}$ of the mode(s) of undistorted cell overlapping with the modes in the distorted cell are also
    reported, with the overlap percentage between parentheses.
  Only peaks identified as one phonon modes either in the experimental references or in this work are listed.}
\label{tab:Ram_lowT}
\begin{tabular}{c c c c c c c ccccc}
\hline
Symmetry & \multicolumn{2}{c}{Exp.}& Comment &  \multicolumn{3}{c}{Calc.}  & {\bf q}& \multicolumn{1}{c}{$\omega_{\rm undist.}$}     \\
         &  53 K \cite{Holy} &  11K \cite{Sugai}&  &HSE(17,0) &  HSE06 & PBE  &      & HSE06     \\ \hline
E$_g$       &        &                        &  &30& 31 &  28   & A & 28.3 (100)     \\              
E$_g$      &   74               &  78  & CDW  &80.3&    72.6    & 55.9   & L  & -126 (95)     \\    
E$_g$       &   93               &  94           & &  94.4  &   97.0   &    93.7 & L & 94.7 (97)       \\    
A$_{1g}$  &   116         &  119      & CDW  & 126&   114   &    96.8 &  L &    -126 (75) + 117 (17)        \\ 
E$_g$      &    114              &  115    &  &115 &    118      &   111 & L &  117 (100)     \\      
A$_{1g}$    &                  &    115  &  &114&  119  &   112 & L & 117 (83) + -126 (16)       \\       
E$_g$      &    136              &   138     &   & 136,139& 141,144    &   132,135  &  A+M,$\Gamma+L$  & 133 (41)+152 (41) and 151 (50) +147 (35) \\
E$_g$      &     148           &                &  weak  \\
E$_g$      &               &           &     & 154 & 155    &  144 &  M+A &  150 (53) + 133 (34) \\
E$_g$      &               &           &     &154  & 157    &  148 & $\Gamma$ + L & 147 (54) + 152 (39)     \\
A$_{1g}$      &                        &   151,158,163  & weak \\ 
A$_{1g}$  &     173       &        176 &  &180 & 180  &  166 & M&  177.1 (94)             \\      
E$_g$     &                       &  &  & 179 &        179         &    168    & M+L+A &  177 (72) + 205 (15) + 133 (10)  \\ 
A$_{1g}$  &    187             &    195   &   &200&  202  &   187 &  M & 204 (93)   \\   
E$_g$     &                            &                        &    & 201  & 203 &  189 & M &  204 (83) + 205 (12)      \\
A$_{1g}$ &     204            &     206    &   &215&  214  &  199 &    $\Gamma$  &  210 (92)   \\
E$_g$    &                  &     212  &  &    217 &     215           & 202  &L$(80\%)$ +M  &  207 (70) + 207 (12)           \\
E$_g$    &        314                     &317   &   &  327  &    328           & 297  & L   &   320 (99)     \\
\hline
\end{tabular}
\end{sidewaystable}

\begin{table}[h!]
  \caption{Infrared phonons at obtained from reflectivity (ref.) and transmission (tr.) spectra.
    Peaks identified in infrared spectra of the CDW phase obtained from reflectivity (ref.) and transmission (tr.)
    experiments compared to computed infrared active frequencies of E$_u$ symmetry.
    The point(s) {\bf q} in the Brillouin zone of the mode(s) of undistorted cell overlapping with the modes in the distorted
    cell are also reported.}
  \label{tab:IRlowT}
\begin{tabular}{ccccc c c c cc}
\hline
\multicolumn{2}{c}{Exp. ref.} &  \multicolumn{2}{c}{Exp. tr.} & Comment & \multicolumn{3}{c}{Calc.}   & {\bf q}   \\
18 K\cite{Liang}& 20 K\cite{Holy} & 28 K \cite{uchida1981IR_Ram} & 4 K  \cite{wilson1978IR_rev} & &HSE(17,0) & HSE06 &   PBE  \\ \hline
      & 42 & 40 & 37 & weak & \\
      & 52 & 47 & 49 & weak & \\
      & 64 & 64 & 64 & weak & \\
      & 76     & 74  & 72  & CDW &73.3 &67.5 &  66              & M   \\
      & 90  & 88  & 100 & weak & 102&105 &  100            & M  \\
118.2 & 118 & 118 & 118 & weak &119 &123&  115              & M  \\
139.5 & 137    & 139 & 140 &  & 137,138&141,143 &  132,137     & $\Gamma$+L,A+M \\
151.3 & 152    & 152 & 153 &  &152,155 &154,156 &  143,150     & M+A,L+$\Gamma$   \\
      & 162  & 166 & 164 & weak &                   &                 \\
171.9 & 175    & 176 & 173 &  &176& 177 &  167            & L+M   \\ 
      & 178    &  & 186 & weak  &                   &                 \\       
196.9 & 198    & 198 & 198 &  &203& 205 &  192            & L+M   \\        
213.0 &        & 214 &211&  &219  &217 & 205            & M+L \\
      &        &     &222&  &  \\
      &        & 230 & 234 & &  &             &  &  \\
      &        & 280 &  &   & &          & &  &  \\
      &        & 315 &  &  &330& 333 &  305          & M  \\
\hline
\end{tabular}
\end{table}

\begin{table}[h!]
  \caption{Computed infrared active frequencies with the
    point(s) {\bf q} in the Brillouin zone and the frequency(s)
    $\omega_{\rm undist.}$ of the mode(s) of undistorted cell overlapping with the modes in the distorted cell.
  The overlap percentage is reported between parentheses.}
  \label{tab:IRdist_und}  
\begin{tabular}{c c c c cccccccc}
\hline
  Symmetry     & HSE(17,0) &PBE & HSE06 & {\bf q} & $\omega_{\rm undist.}$  HSE06      \\ \hline
    E$_u$   & 73.3 & 66.1 & 78.9  & M  & -100 (96)  \\ 
    A$_{2u}$  & 101 & 99.7 & 105  & M  &  102 (92) \\ 
    E$_u$   & 102 &100 & 106  & M  &  102 (96) \\ 
    E$_u$   & 119 &115   & 123  & M  & 122 (99)\\ 
    E$_u$   & 137 &132   & 141  & $\Gamma$+L  & 124 (60) + 152 (30)  \\ 
    E$_u$   & 138 &136   & 143  & A+M  & 149 (60) + 147 (30) \\          
    A$_{2u}$  & 143 &137   & 150  & L+M  & 152 (74) + 147 (25) \\ 
    A$_{2u}$  & 146 &139   & 151  & M+L  & 147 (71) + 152 (26) \\  
    E$_u$   & 152 &143   & 155  & M+A  & 147 (62) + 149 (33) \\
    E$_u$   & 155 &150   & 158  & L+$\Gamma$  & 152 (68) +124 (24) \\ 
    E$_u$   & 176 &167   & 177  & L+M  & 176 (75) +  207 (12) \\
    E$_u$   & 203 &192   & 206  & L+M  & 207 (78) + 207 (16)  \\ 
    A$_{2u}$  & 202 &197   & 208  & M & 207 (98)  \\ 
    E$_u$   & 219 &205   & 218  & M+L & 207 (76) + 207 (15) \\
    A$_{2u}$  & 320 &296   & 328  & $\Gamma+$M  & 323 (69) + 329 (30)  \\ 
    A$_{2u}$  & 326 &301   & 332  & M$+\Gamma$  & 329 (69) + 323 (30) \\ 
    E$_u$   & 330 &305   & 334  & M & 329 (99) \\ 
  \hline
\end{tabular}
\end{table}
\FloatBarrier
\section{CDW instability as a function of U}
In Fig. \ref{res} we present the energy gain in the CDW distortion obtained with the PBE+U method for 
values of U between 0 to 2 eV. We see that the energy gain reduces monotonically with U and vanishes 
at 1.5 eV. This can be compared to the LDA+U result of U=2.5 eV in Ref. \onlinecite{bianco2015}. The PBE+U 
band-structure agrees well with HSE06 at U=3.25 eV. The corresponding value in LDA+U is 3.5 eV.
\begin{figure}[t]
\includegraphics[scale=0.42]{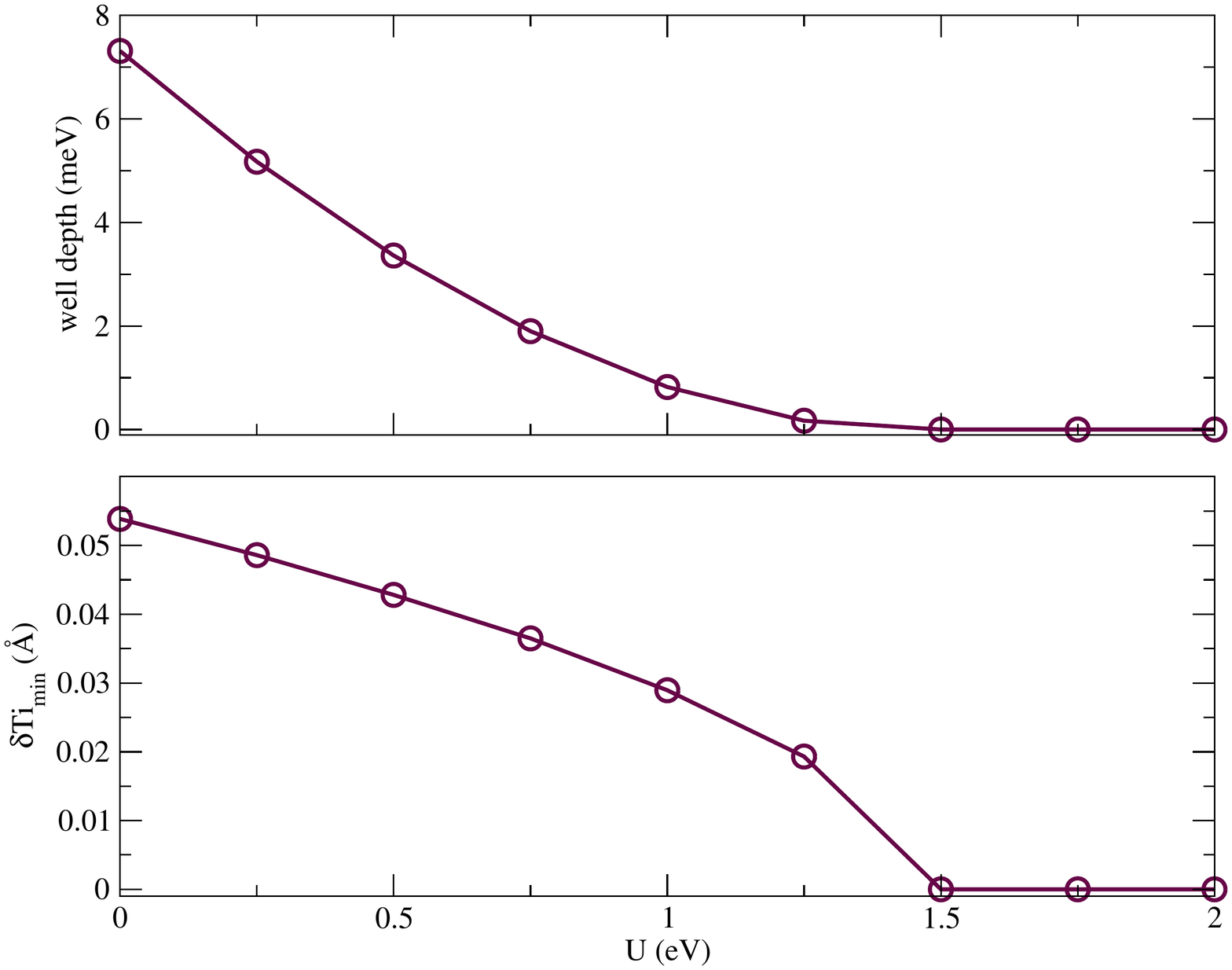}
\caption{The upper panel shows the absolute energy gain or well-depth as a function of U within PBE+U and the 
lower panel shows the corresponding displacement $\delta {\rm Ti}_{\rm min}$.}
\label{res}
\end{figure}
\section{Spin-orbit coupling}
The calculations presented in the main paper were obtained without spin-orbit coupling (SOC). We did, however,
also perform calculations including SOC (using VASP since SOC is not an available option in {\tt CRYSTAL}). We 
verified that SOC only added a minor quantitative correction and did not change the qualitative analysis. 
Including SOC increases the computational cost substantially and we could therefore only carry out a limited 
number of calculations. Furthermore, to make the calculations feasible we had to omit the $3p$ semi-core 
electrons in the Ti pseudo potential. This increases the energy gain but is not expected to change the 
effect of SOC, since SOC influences mainly the Se bands. The semi-core 
electrons are also not expected to largely influence the electron-phonon coupling (EPC) of the high-T phase. 
Indeed, we found a difference in the EPC of only 4\%. In Fig.~\ref{socnosoc} we have summarised 
our analysis using HSE06. At large displacement $\delta$Ti, the energy gain is around twice as large when 
omitting semi-core electrons. The effect is smaller, the smaller is the distortion, consistent with the fact that 
both the band structure and the EPC are very similar with and without semi-core electrons. 
   
Including SOC we find only a very small change, the maximum being 2 meV at the minimum of the curve. 
At small distortion the difference is vanishingly small. The effect of SOC reduces the EPC in HSE06 from 
$19 \,{\rm eV}^2/{\rm \AA}^2$ to $16 \,{\rm eV}^2/{\rm \AA}^2$, which is compensated by the change in 
band structure. Apart from the indirect effect of SOC on the orbitals and the self-consistent potential 
there is a direct effect due to extra SOC terms in the Hamiltonian.\cite{soc} Since these terms are semi-local and 
density independent, we expect them to give only a small contribution to the non-local $p-d$ 
EPC and to be the same for all HSE functionals having the same band structure. The indirect effect is also 
expected to be small and similar for all 
HSE functionals. Indeed, similarly to HSE06, a reduction of approximately $3 \,{\rm eV}^2/{\rm \AA}^2$ 
is found for HSE(30,0.3), which is accompanied by an identical change in band structure, and no 
difference in energy gain. We therefore expect that the relative EPC strengths with SOC and non local functionals 
are qualitatively the same as those found without SOC.
\begin{figure}[t]
\includegraphics[scale=0.45]{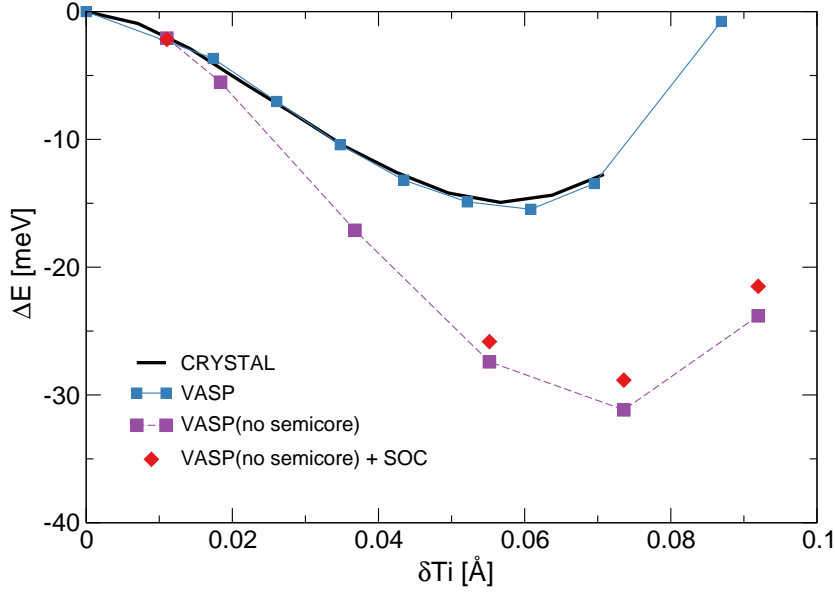}
\caption{Energy gain within HSE06 calculated with {\tt CRYSTAL} and VASP. We compare results with and without 
semi-core electrons in the Ti-PAW potential, and with and without SOC. }\label{socnosoc}
\end{figure}
\section{Electronic structure compared to ARPES}
In order to compare the band structure with experiment it is important to take into account
the Se-$p$ band splitting around $\Gamma$ in the high-T phase which is due to SOC. \cite{vydrova2015} 
In Fig. \ref{bands} we report calculated band structures in 
the high (a) -and low-T (b) phases superimposed on ARPES measurements by Rohwer et al.. \cite{rohwer}
In the high-T phase we calculated the bands 
along $M'-\Gamma-M$ (black), $L'-\Gamma-L$ (orange) and along a line in between with $k_z=0.33$ (red). 
The band dispersions are well-reproduced and the band overlap ($\Gamma-L$) is only slightly overestimated 
with HSE06. The bands in the CDW phase are calculated at the minimum of the curve in Fig. 1b 
of the manuscript and we note that band gap is indirect between $\Gamma$ and $A$. The gaps in 
HSE06 and HSE(17,0) are 0.21 eV and 0.35 eV, respectively. Both are somewhat overestimated compared to 
the experimental value which has been estimated to 0.15 eV.

 \begin{figure}[b]
\includegraphics[scale=0.5]{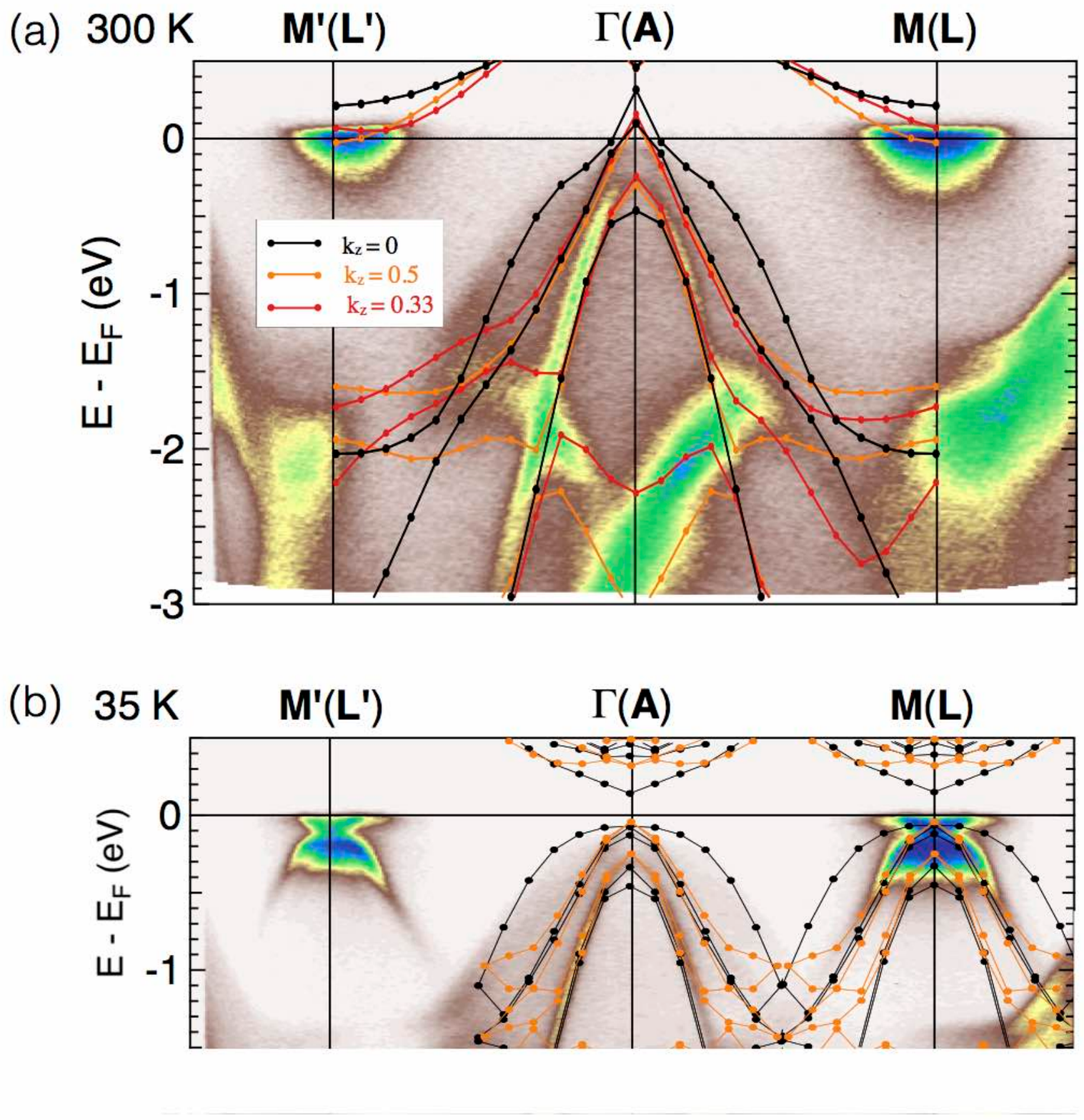}
\caption{HSE06 electronic structure of the high (a) and low-T phases (b) compared to ARPES. $k_z$ values are given 
in units of $2\pi/c$.}\label{bands}
\end{figure}
\newpage

\end{widetext}
\end{document}